\documentclass[twocolumn]{aastex631}
\usepackage{multirow}
\usepackage{amsmath}
\newcommand{\qdfit}{\texttt{q3dfit}}

\newcommand{\um}{$\mu$m}

\newcommand{\oiii}{[\textrm{O}~\textsc{iii}]}

\newcommand{\ha}{${\rm H\alpha}$}
\newcommand{\hb}{${\rm H\beta}$}

\newcommand{\jwst}{JWST}

\shorttitle{Q3D: Host galaxy}
\shortauthors{Chen et al.}

\begin{document}

\title{JWST IFU observations uncover host galaxy continua in extremely red and obscured quasars}

\newcommand{\jhu}{Department of Physics and Astronomy, Bloomberg Center, Johns Hopkins University, Baltimore, MD 21218, USA}
\newcommand{\stsci}{Space Telescope Science Institute, 3700 San Martin Drive, Baltimore, MD 21218, USA}
\newcommand{\ias}{Institute for Advanced Study, Princeton University, Princeton, NJ 08544, USA}
\newcommand{\uiuc}{Department of Astronomy, University of Illinois at Urbana-Champaign, Urbana, IL 61801, USA}

\correspondingauthor{Yu-Ching Chen}
\email{ycchen@jhu.edu}

\author[0000-0002-9932-1298]{Yu-Ching Chen}
\affiliation{\jhu}

\author[0000-0001-6100-6869]{Nadia L. Zakamska}
\affiliation{\jhu}

\author[0000-0002-0710-3729]{Andrey Vayner}
\affiliation{\jhu}
\affiliation{\rm IPAC, California Institute of Technology, 1200 E. California Boulevard, Pasadena, CA 91125, USA}

\author[0000-0001-7351-2531]{Jack M. M. Neustadt}
\affiliation{\jhu}

\author[0000-0003-2212-6045]{Dominika Wylezalek}
\affiliation{Zentrum für Astronomie der Universität Heidelberg, Astronomisches Rechen-Institut, Mönchhofstr 12-14, D-69120 Heidelberg, Germany}

\author[0000-0002-1608-7564]{David S. N. Rupke}
\affiliation{Department of Physics, Rhodes College, 2000 N. Parkway, Memphis, TN 38112, USA}

\author[0000-0002-3158-6820]{Sylvain Veilleux}
\affiliation{Department of Astronomy and Joint Space-Science Institute, University of Maryland, College Park, MD 20742, USA}

\author[0000-0002-6948-1485]{Caroline Bertemes}
\affiliation{Zentrum für Astronomie der Universität Heidelberg, Astronomisches Rechen-Institut, Mönchhofstr 12-14, D-69120 Heidelberg, Germany}

\author[0000-0001-7572-5231]{Yuzo Ishikawa}
\affiliation{MIT Kavli Institute for Astrophysics and Space Research, Massachusetts Institute of Technology, Cambridge, MA 02139, USA}

\author[0000-0001-9755-9406]{Marie Wingyee Lau}
\affiliation{Department of Physics and Astronomy, University of California, Riverside, CA 92521, USA}

\author[0000-0003-3762-7344]{Weizhe Liu}
\affiliation{Department of Astronomy, Steward Observatory, University of Arizona, Tucson, AZ 85719, USA}

\author[0000-0002-3191-8151]{Marshall D. Perrin}\affiliation{Space Telescope Science Institute, 3700 San Martin Drive, Baltimore, MD 21218, USA}






\begin{abstract}
Uncovering bright quasars' host galaxies at cosmic noon is challenging because of the high contrast between the quasar and its host and redshifted light, making them primarily visible in the infrared.
We present JWST NIRSpec integral field unit (IFU) observations of six extremely red quasars (ERQs) at $z=2.4-2.9$ and two dust-obscured quasars at lower redshifts. Using image decomposition across the spectral range, we successfully separate quasar and host galaxy continuum emission, model host morphologies, and extract spectra. 
The ERQs and obscured quasars have compact host galaxies with half-light radii of 1.4$-$2.9 kpc and stellar masses of 10$^{10.6-10.9}$ $M_{\odot}$. Their stellar masses are consistent with the average stellar mass of quasar hosts as expected from abundance matching and clustering analysis.
Most of the quasars in our sample exhibit significant spatial offsets (0.4$-$1.3 kpc) between the quasar and host galaxy, potentially caused by post-merger dynamics or non-uniform dust obscuration. The ERQs reside 0.5$-$2 dex above the local black hole-stellar mass relation, similar to other heavily obscured populations such as HotDOGs, optically selected quasars at cosmic noon, and high-redshift SMBH candidates identified with JWST. However, this ``over-massive" feature might be attributed to selection bias. 
Compared to HST-based studies, our JWST measurements reveal more compact host galaxies, smaller Sersic indices, and lower stellar masses, likely because of improved resolution, more accurate modeling, and minimal line contamination. These findings highlight the unique capabilities of JWST IFU in revealing quasar host galaxy properties and potential evolutionary stages of obscured quasars at cosmic noon.
\end{abstract}



\section{Introduction} \label{sec:intro}

The shape of the galaxy luminosity function and the reduced star formation in elliptical galaxies suggest that feedback from active supermassive black holes, known as quasars, plays a critical role in suppressing star formation \citep{Silk1998,Croton2006,Fabian2012}. Additionally, the tight correlation between supermassive black holes and the properties of their host galaxies implies a potential co-evolution between them \citep{KormendyHo2013,McConnell2013}.
A widely accepted evolutionary paradigm, supported by cosmological simulations, proposes that galaxy mergers compress molecular gas, trigger enhanced star formation and funnel gas into the central supermassive black holes \citep{Sanders1988,Hopkins2006}. This accretion process begins in an obscured phase; as quasar feedback intensifies and expels surrounding dust and gas, the quasar becomes visible in the optical. 

Obscured quasars thus provide excellent targets to test this evolutionary sequence. Samples such as hot dust-obscured galaxies (HotDOGs) and extremely red quasars (ERQs) represent distinct categories of dust-obscured quasars \citep{Eisenhardt2012,Wu2012,Ross2015,Hamann2017}. 
ERQs are selected based on their very red colors in the rest-frame UV to near-infrared, typically using data from surveys such as the Sloan Digital Sky Survey (SDSS) in the optical and the Wide-field Infrared Survey Explorer (WISE) in the mid-infrared \citep{Ross2015,Hamann2017}. In contrast, HotDOGs are primarily identified from the WISE survey using the ``W1W2-dropout" technique, that is, these sources are faint or undetected in the WISE W1 (3.4 $\mu$m) and W2 (4.6 $\mu$m) bands \citep{Eisenhardt2012,Wu2012}.
ERQs, in particular, are thought to display the extreme quasar feedback, characterized by extended, fast-moving outflows traced by ionized [O III] gas \citep{Zakamska2016_erq,Perrotta2019,Vayner2021a}.

Because the emission from the central quasar is significantly brighter than that of its host galaxy, and the angular diameter distance increases substantially at $z>0.1$, detecting host galaxies at high redshift requires high angular-resolution facilities with exceptional sensitivity to disentangle the host galaxy from the quasar. The integral field unit (IFU) capability of the newly launched James Webb Space Telescope (JWST) in the near-infrared wavelength opens a new window for studying quasar feedback in detail, both spatially and spectrally \citep{Gardner2006,Jakobsen2022}.
Several studies using JWST IFU have revealed ionized cones, clumpy star formation, and shocks in host galaxies, all associated with powerful outflows \citep{Wylezalek2022,Cresci2023,Veilleux2023,Vayner2023,Vayner2024,LiuW2025}. However, the stellar continua of these obscured quasars have not yet been explored with JWST data. 
Recent studies using near-infrared imaging from the Hubble Space Telescope (HST) have successfully revealed the stellar light from the host galaxies of obscured quasars \citep{Glikman2015, Fan2016, Farrah2017, Zakamska2019}. However, the connection between these obscured quasars and galaxy mergers remains debated -- while some studies report a high merger fraction \citep{Glikman2015, Fan2016}, others find no significant evidence for elevated major merger activity \citep{Farrah2017, Zakamska2019}.

In this paper, we present the stellar light from the host galaxies of eight extremely red and obscured quasars at redshift $z\sim2.4$, as well as at $z=$ 0.4, 1.6 and 2.9. In Section \ref{sec:data}, we describe the target selection, observation details, and data reduction and analysis methods. Our results, including wavelength-averaged images and host galaxy spectra, are presented in Section \ref{sec:results}. In Section \ref{sec:discussion}, we explore the physical properties of the host galaxies, the reasons behind the spatial offsets between quasars and their hosts, the black hole mass-stellar mass relation, and compare our findings with previous studies. Finally, we summarize our findings in Section \ref{sec:conclusion}. We adopt a flat $\Lambda$CDM cosmology with $\Omega_\Lambda=0.7$, $\Omega_m=0.3$, and $H_0=70\,{\rm km\,s^{-1}Mpc^{-1}}$ throughout this paper. 

\section{Observations, Data reduction and analysis} \label{sec:data}

\subsection{Target selection}

Our targets include near-Eddington extremely red quasars at $z\sim2.5$  (ID: 2457, PI: Vayner) and obscured quasars with known outflows from local universe to $z\sim3$ from the early release program (ID: 1335, PI: Wylezalek; Co- PIs: Veilleux, Zakamska; Software Lead: Rupke)
The five extremely red quasars from the first program (ID: 2457) are a subset of quasars with extremely red infrared-to-optical colors and large C IV rest equivalent widths \citep{Ross2015,Hamann2017}. They are also exhibit very broad and strongly blueshifted [O III] $\lambda$5007 emission lines \citep{Zakamska2016_erq,Perrotta2019}. Those features are thought to be associated with exceptionally powerful outflows from dust-obscured quasars, which are still in the early stages of massive
galaxy evolution \citep{hopkins08}. 
The three targets from the early release program \citep[ID: 1335;][]{Wylezalek2022} are powerful, dust obscured quasars with known high-velocity [O III] outflows. The three targets are F2M1106 at $z=0.4350$ \citep{ShenL2023}, XID2028 at $z=1.5927$ \citep{Perna2015}, and J1652+1728 at $z=2.9548$ \citep{Gillette2024}. Both F2M1106 and XID2028 show spatially resolved [O III] outflows in the integral-field-unit data \citep{ShenL2023,Cresci2015}, and J1652+1728 is part of the extremely red quasar sample \citep{Ross2015,Hamann2017}.

\subsection{JWST observations}
The eight targets were observed with James Webb Space Telescope (JWST) in the integral field unit (IFU) mode of the Near Infrared Spectrograph \citep[NIRSpec;][]{Boker2022} between November 2022 and June 2024. The field-of-view of the JWST IFU data is $3 \arcsec\times3\arcsec$. The G235H/F170LP grating (1.66–3.17 $\mu$m) was used for all targets except XID2028. XID2028 was observed only with the G140H/F100LP grating (0.97–1.89 $\mu$m). F2M1106 was also observed with the G395H/F290LP grating (2.87–5.27 $\mu$m). These high-resolution gratings provide spectral resolutions of $R\sim$ 1500-3500 ($\Delta v\sim$ 85-200 km/s), which are excellent for modeling gas and stellar dynamics. We obtained a leakage exposure with the same grating in one dither position for each target to remove the light from the failed-open shutters and the light leaking through the micro-shutter assembly (MSA) \citep{Deshpande2018}. We used a nine-point small cycling dither pattern for the three targets in program 1335, and a ten- or eleven-point small cycling dither pattern for the five targets in program 2457 to improve spatial sampling and help us more accurately measure and characterize the point-spread function (PSF). The observation details are listed in \autoref{tab:obs}.
All the JWST data used in this paper can be found in MAST: \dataset[10.17909/r93w-tj43]{http://dx.doi.org/10.17909/r93w-tj43}.

\begin{deluxetable*}{cccccccc}
 \tablecaption{Observation details of eight extremely red and obscured quasars.
 \label{tab:obs}}
 \tablehead{ \colhead{Abbreviated Name} & \colhead{R.A.} & \colhead{Decl.} & \colhead{$z$}  &  \colhead{Grating} & \colhead{Exp. Time}  & \colhead{Obs. Date} &\colhead{Program ID} \\ 
 \colhead{(J2000)} &  \colhead{(degree)} &\colhead{(degree)} & & & \colhead{(hour)} & \colhead{(UT)}& } 
 \colnumbers
 \startdata
SDSSJ0832+1615 & 128.000936 & 16.25006 & 2.4249 & G235H/F170LP & 2.23 & 2022-12-09 & \multirow{5}{*}{2457} \\
SDSSJ0834+0159 & 128.702086 & 1.98916 & 2.5850 & G235H/F170LP & 2.03 & 2022-11-13 & \\
SDSSJ1217+0234 & 184.269542 & 2.57146 & 2.4280 & G235H/F170LP & 2.03 & 2023-06-21 \\
SDSSJ1232+0912 & 188.173885 & 9.20259 & 2.4050 & G235H/F170LP & 2.23 & 2024-06-20\\
SDSSJ2215$-$0056 & 333.849987 & -0.94546 & 2.5093 & G235H/F170LP & 2.23& 2022-11-23\\
\hline
\multirow{2}{*}{F2M1106} & \multirow{2}{*}{166.701481} & \multirow{2}{*}{48.12002} & \multirow{2}{*}{0.4350} & G235H/F170LP & 0.55 & 2022-11-13 & \multirow{4}{*}{1335}\\
& & & & G395H/F290LP & 0.55 & 2022-11-13 \\
XID2028 & 150.547124 & 1.61847 & 1.5927 & G140H/F100LP& 2.92 & 2022-11-20  \\
SDSSJ1652+1728 & 253.011006 & 17.48128 & 2.9548 & G235H/F170LP & 4.56 & 2022-07-15
\enddata
 \tablecomments{Column 1: Target name. Column 2: Right Ascension. Column 3: Declination. Column 4: Redshift from the literature \citep{Perna2015,ShenL2023,Gillette2024}. Column 5: Grating. Column 6: Exposure Time. Column 7: Observation Date. Column 8: Program ID.}
\end{deluxetable*}

\subsection{NIRSpec data reduction}
\label{sec:data_reduction}

We process the data using \jwst\ calibration pipeline version 1.14.0 \citep{Bushouse2023.1.14.0}. First, we run the \texttt{Detector1} pipeline, which applies detector-level corrections to all uncalibrated science exposures and converts raw images into corrected count rate files. The steps include various corrections such as bad pixel masking, bias correction, dark current correction. Given that the NIRspec‘s detectors are readout noise limited, we employ an additional routine using the \texttt{NSClean} algorithm to remove the picture frame pattern \citep{Rauscher2024_NSClean}. After the correlated read noises are removed, the count rate files are processed with the \texttt{Spec2} pipeline. During the \texttt{Spec2} pipeline, various instrument-specific calibrations are performed including coordinate assignment, flat-field correction, flux calibration, background subtraction, and so on. 

\begin{figure}
  \centering
    \includegraphics[width=\columnwidth]{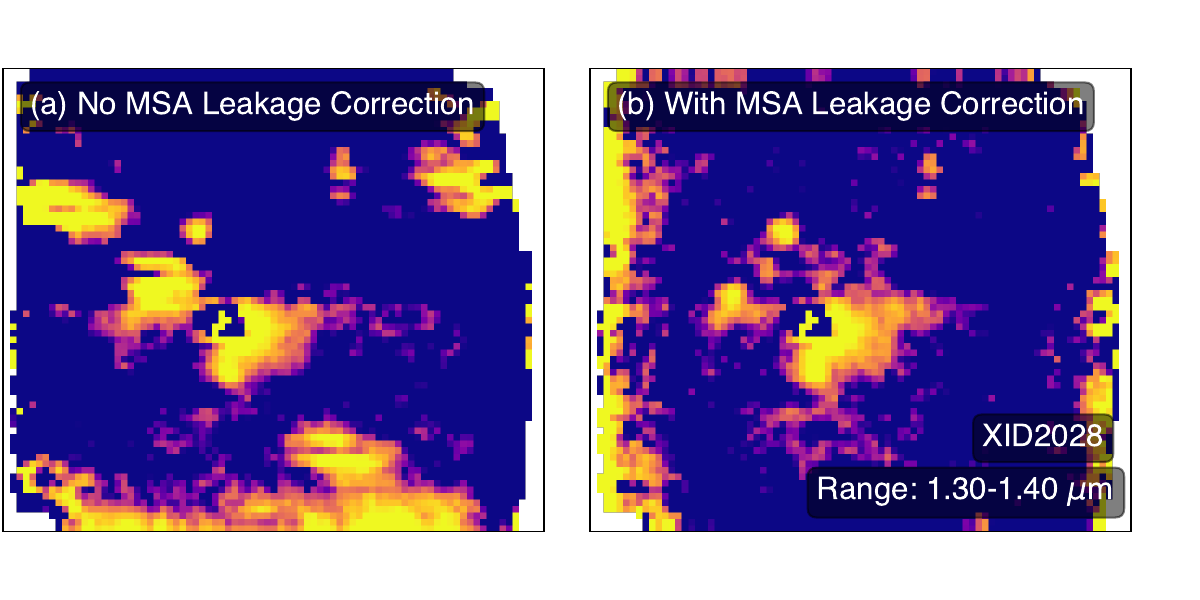}
    \caption{Example wavelength-averaged PSF-subtracted images before and after MSA leakage correction. Several bright stripe-like artifacts disappear after the MSA leakage correction.
    }
    \label{fig:leakcal_comparison}
\end{figure}

Removing MSA leakage artifacts, particularly those from failed-open shutters, is crucial for accurately modeling the stellar continuum. These artifacts appear as bright, featureless sources at fixed positions across all wavelengths. We apply MSA leakage correction using a dedicated leakage exposure taken at one of the dither positions. While this method introduces additional noise, about three times higher in our case due to using one leakcal exposure and nine/ten dithers, it effectively removes artifact light caused by failed-open shutters and MSA print-through. Without correction, this contamination could be mistaken for genuine astrophysical sources (\autoref{fig:leakcal_comparison}). In the future, we plan to generate a leakage mask for failed-open shutters, which would eliminate the artifacts without adding background noise.

To remove the low-level cosmic rays and bad pixels in the \texttt{cals.fits} files, we conduct an additional 2.5$\sigma$ rolling outlier rejection with a window size of 10 pixels along the wavelength direction (x-direction) in each row (spaxel). We then combine the fully calibrated individual exposure at each dither position into a final 3-dimensional drizzled data cubes using the standard \texttt{Spec3} pipeline. We choose \texttt{drizzle} algorithm instead of the Shepard’s method of weighting \texttt{emsm} algorithm to avoid further smoothing along the spectral direction \citep{Law2023}. The final pixel size is set to be 0\farcs05$\times$0\farcs05, half of the default pixel size, to better capture the structure of JWST's point spread function (PSF). The wavelength sampling is 2\AA, 3\AA, and 5\AA\ for G140H/F100LP, G235H/F170LP, and G395H/F290LP gratings, respectively, which corresponds to approximately three times the spectral resolution for each setup.

\subsection{PSF subtraction} \label{sec:psf_subtraction}

Detecting the subtle emission from the host galaxy of the luminous quasar often require  specialized techniques, such as dedicated Point Spread Function (PSF) subtraction and spectral decomposition methods. 
In previous work from the ERS-1335 program, we used the dedicated software \qdfit\ to subtract the PSF \citep{Rupke2023_q3dfit}. \qdfit\ performs spectral decomposition on each spaxel in the cube, guided by a spectral prior derived from the quasar spectrum extracted from the brightest spaxel. This method has proven highly effective in revealing faint extended ionized gas emission, even when the quasar contributes to the same spectral features, due to the distinct velocity differences between nuclear and extended gas components \citep{Wylezalek2022, Vayner2023, Veilleux2023, Vayner2024,Rupke2017}. 
\qdfit\ also includes  functionality to identify and characterize quasar host emission via spectral decomposition. Spectral decomposition has been used to detect the stellar component, including stellar kinematics, in quasar hosts in the local universe \citep{Rupke2017}.
However, the complex wavelength-dependent structure of the JWST PSF and undersampling in IFU data often introduce spectral ``wiggles" that vary across spaxels. In our JWST data, we find that the stellar continuum is frequently degenerate with both the nuclear continuum and the additive Legendre polynomials used by \qdfit\ to correct for these artifacts and backgrounds.

Here, we introduce a new PSF subtraction technique (Chen et al., in prep.) that performs spatial decomposition using dedicated standard star observations (ID: 3399; PI: Perrin). The standard star was observed in all three high-resolution gratings with 16 dither positions. We reduced these data using the same procedure as for our science targets and generated data cubes with half the pixel size and identical wavelength sampling. To account for the wavelength-dependent PSF, we construct a two-dimensional PSF model at each wavelength using the standard star data. Quasar emission is then removed by fitting and subtracting this wavelength-dependent PSF along the spectral axis using the image decomposition tool \texttt{GALFIT} \citep{Peng2002,Peng2010}. A PSF and a uniform background component are simultaneously fitted during this process. The centroid and flux of the PSF model are treated as free parameters. \texttt{GALFIT} can perform sub-pixel fitting to accurately account for subtle PSF shifts. \texttt{GALFIT} then produces the best-fit model by minimizing the reduced $\chi^2$ in the residual images.


\subsection{Fitting host morphologies and extracting spectra} \label{sec:galfit_specfit}

To obtain the host galaxy morphology and extract its spectrum, we first analyze wavelength-averaged, PSF-subtracted images to determine the galaxy's structural properties, then perform forced photometry across wavelengths to obtain a one-dimensional spectrum. We generate wavelength-averaged images using spectral ranges free of strong emission lines to ensure uncontaminated measurements of the extended continuum emission. We model the host galaxy using a single Sersic profile \citep{Sersic1968}, excluding the central 2–4 pixel region, depending on the quasar’s brightness, to avoid artifacts from PSF undersampling. For J1217+0234, whose host detection is marginal, we fix the host centroid to that of the quasar. During the fit, we constrain the Sersic index to the range 0.5–8 and the axis ratio to 0.2–1. Finally, using the fitted Sersic model with all parameters fixed except for magnitude, we perform forced photometry across all wavelengths to extract the host galaxy’s one-dimensional spectrum.

\subsection{Full spectrum fitting} \label{sec:ppxf_fit}
We utilize the full spectrum fitting code \texttt{pPXF} \citep{Cappellari2017,Cappellari2023} to extract the stellar kinematics and stellar population of galaxies. \texttt{pPXF} fits a set of stellar population synthesis (SPS) models to the observed spectrum, determining key galaxy properties such as ages, metallicity, and stellar velocity dispersion. For our analysis, we select the GALAXEV SPS model templates \citep{Bruzual2003}, which offer a spectral resolution of 3\AA\ and cover a broad range of ages (10$^5$ to 2$\times$10$^{10}$ years) and metallicities ([M/H] = [-1.74, -0.73, -0.42, 0, 0.47]). In addition to the stellar components, we also simultaneously fit the gas emission lines using Gaussian components with shared radial velocities and velocity dispersions. Given the low signal-to-noise ratio of the spectra, we do not apply multiplicative or additive Legendre polynomials because they typically degenerate with the spectral slope.

\section{Results} \label{sec:results}

\subsection{Wavelength-average images and Host Galaxy Morphology}

\begin{figure*}
  \centering
\includegraphics[width=0.49\textwidth]{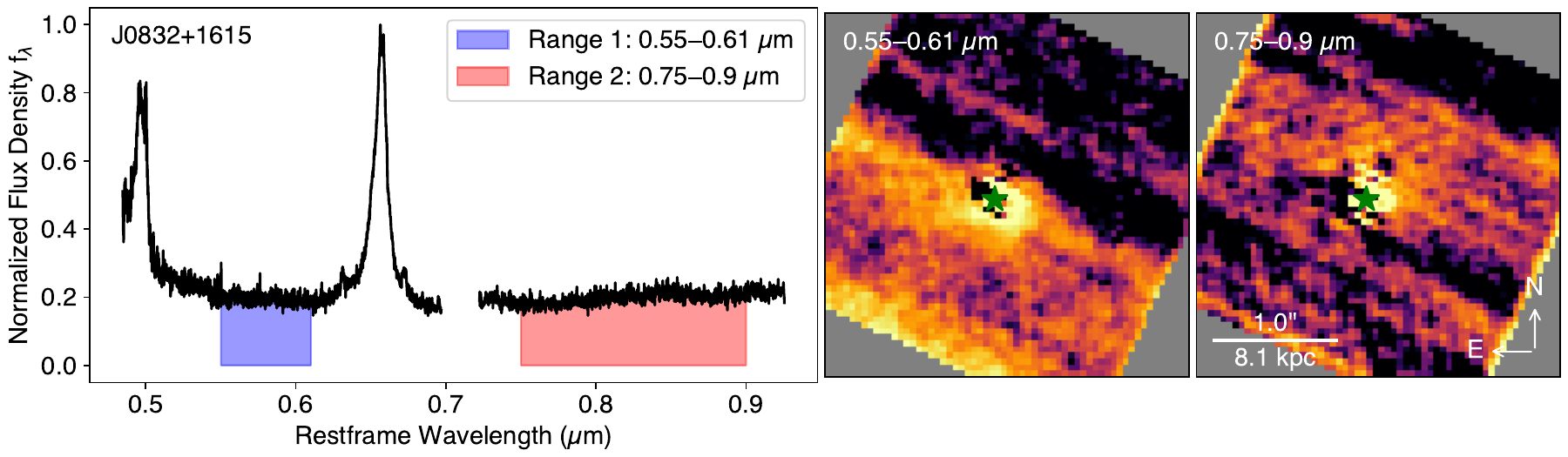}
\includegraphics[width=0.49\textwidth]{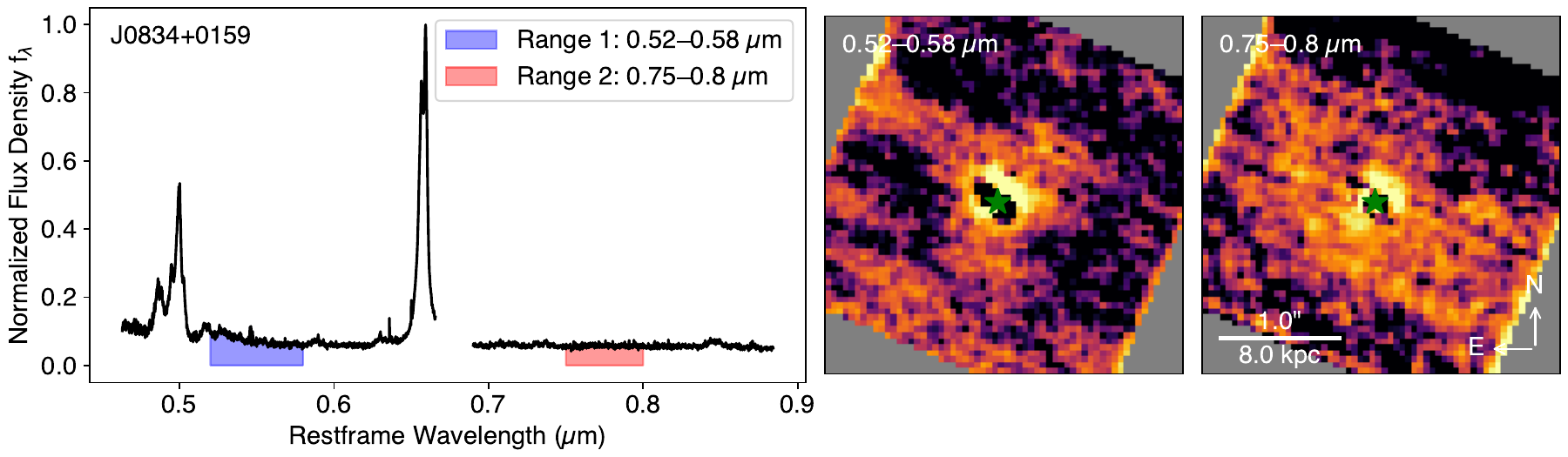}
\includegraphics[width=0.49\textwidth]{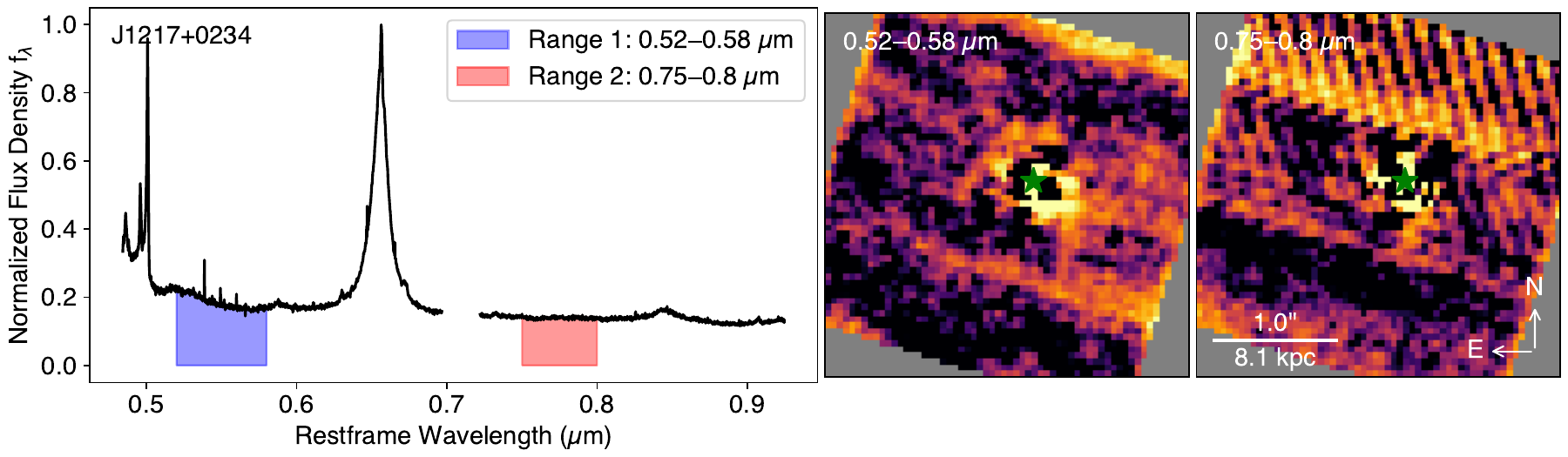}
\includegraphics[width=0.49\textwidth]{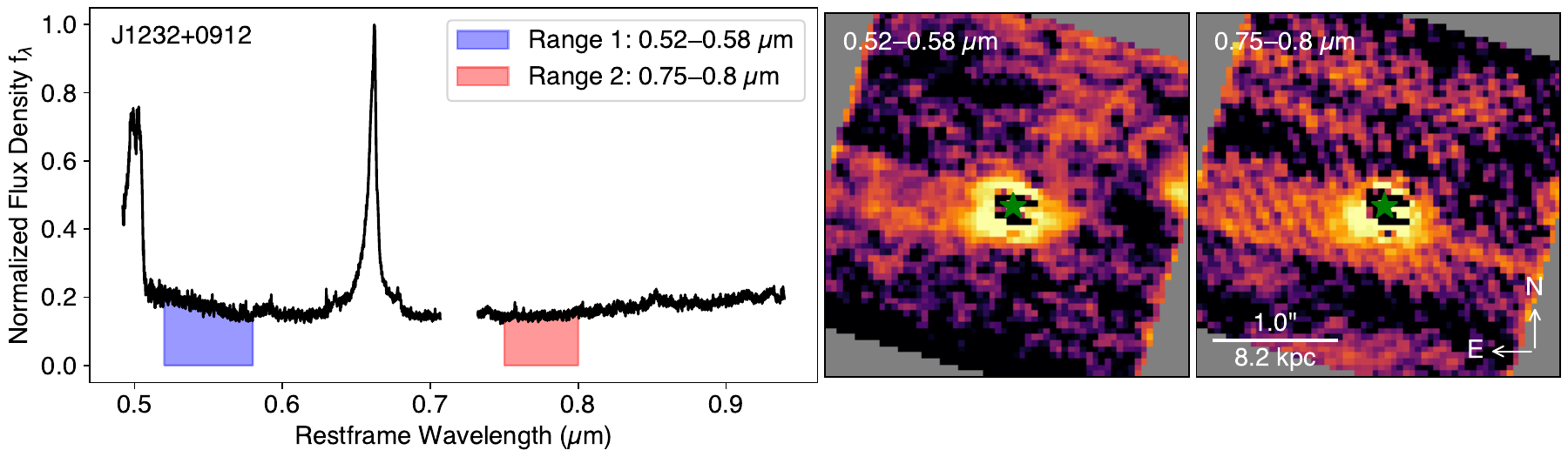}
\includegraphics[width=0.49\textwidth]{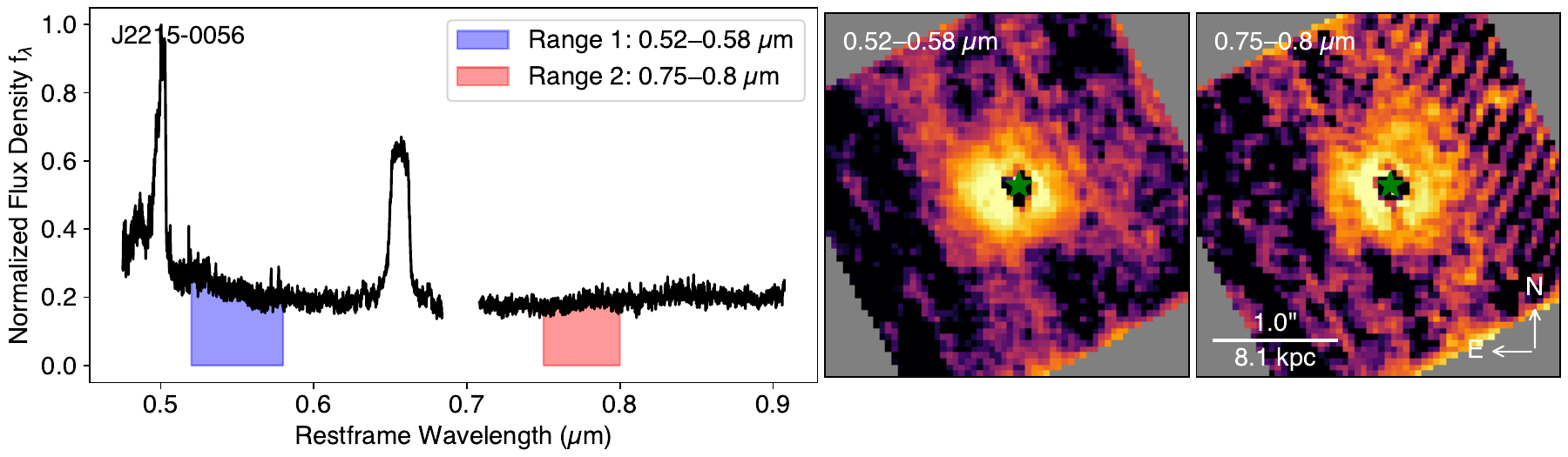}
\includegraphics[width=0.49\textwidth]{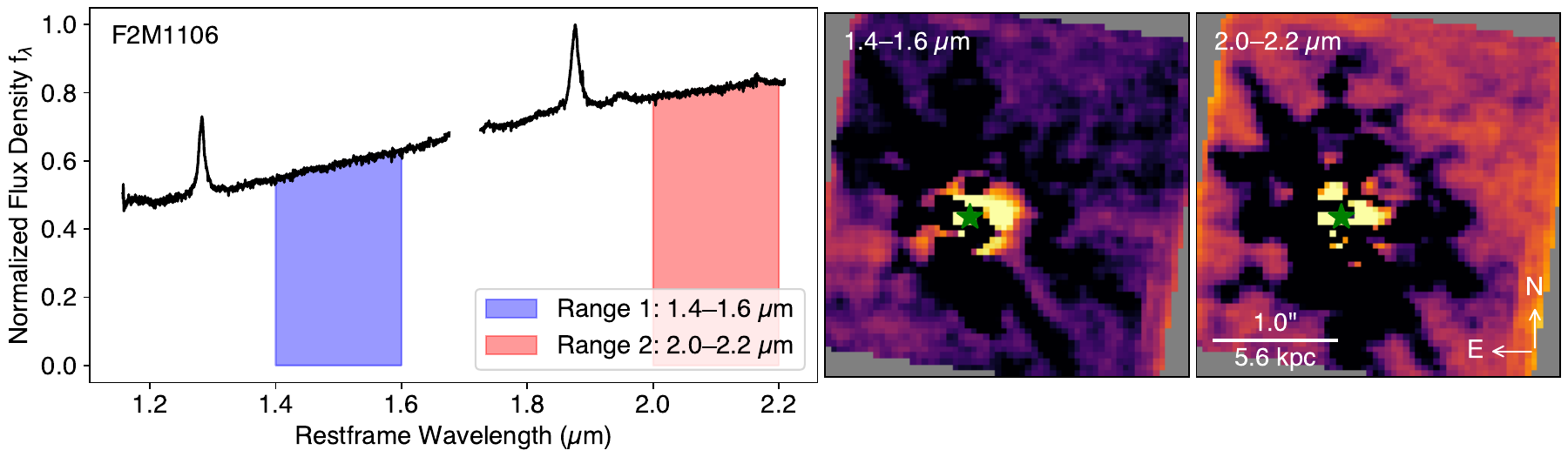}
\includegraphics[width=0.49\textwidth]{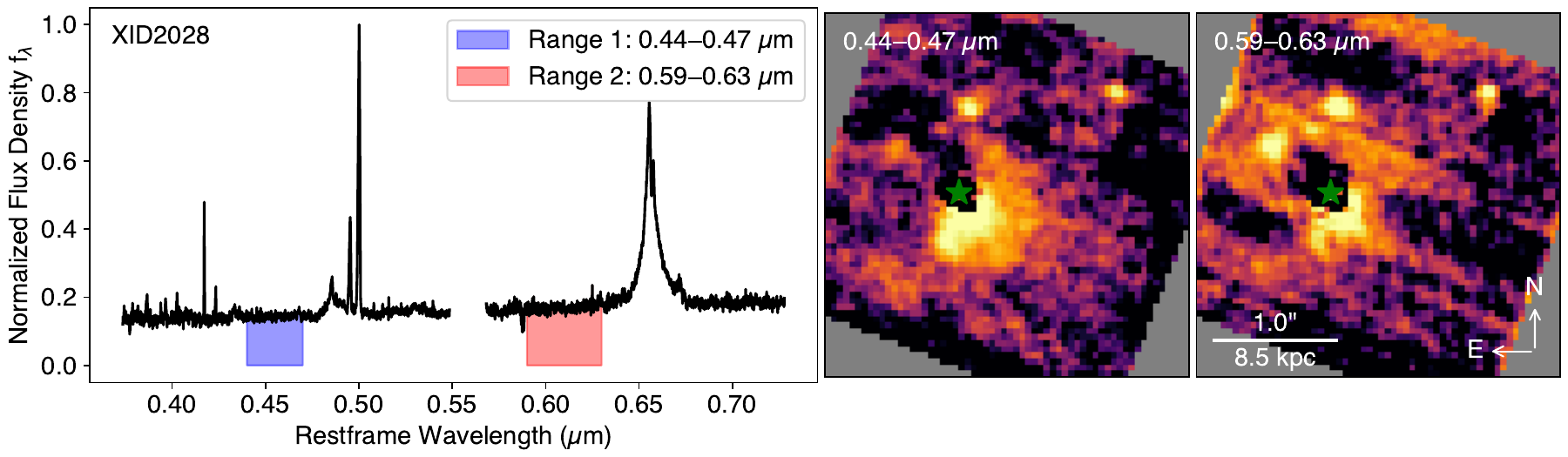}
\includegraphics[width=0.49\textwidth]{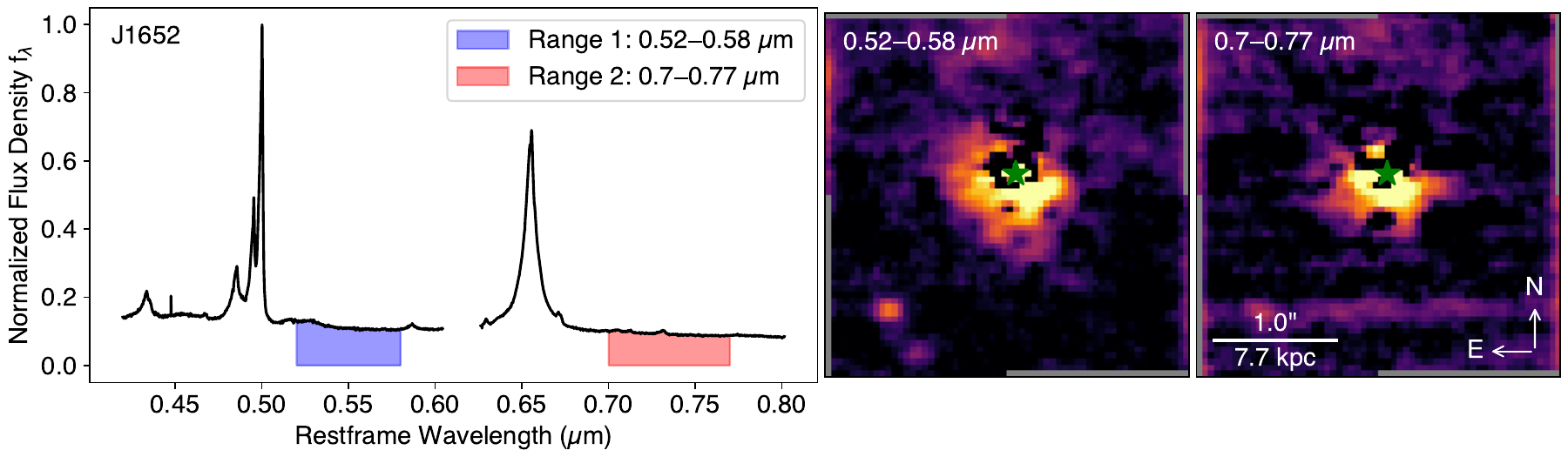}
    \caption{Total spectra and wavelength-average quasar-subtracted images of the eight targets. The left panel of each target shows the total spectrum with shaded regions, which represents the wavelength ranges used for the images. The right two panels show wavelength-average images after PSF subtraction. The green stars represent the quasar locations. The north is up and the east to the left.
    }
    \label{fig:continuum_images}
\end{figure*}

We present the wavelength-averaged, PSF-subtracted images of all eight targets in \autoref{fig:continuum_images}. These broad-band images are created using spectral channels free of strong emission lines, as indicated by the colored regions in the left panels of the figure. We avoid regions with any potential emission lines, even if they were not visible in the integrated spectra, unless the host galaxies are very faint (e.g., J0832+1615). The two regions in each detector were selected and plotted due to their correspondence with distinct detector fields and varying MSA/background patterns, allowing for independent confirmation of the emission's reality. Despite faint striped patterns from residual MSA leakage, we detect extended host galaxy continuum emission in all targets. We noted slight over-subtraction in the central few pixels around the quasars; this is likely due to the PSF attempting to compensate for non-uniform background or noise by minimizing the reduced $\chi^2$. These negative pixels were masked during the subsequent host galaxy fitting. For J0832+1615 and J1217+0234, however, the detections are marginal and approach the noise level. Benefiting from JWST’s high spatial resolution in the near-infrared (full-width-half-maxima (FWHM) $\sim$ 0$\farcs$066 at 2 $\mu$m), we also detect nearby companion galaxies in J1652+1728 and XID2028, previously observed only in ionized or molecular gas \citep{Brusa2018,Vayner2023}.

\begin{figure*}
  \centering
\includegraphics[width=0.49\textwidth]{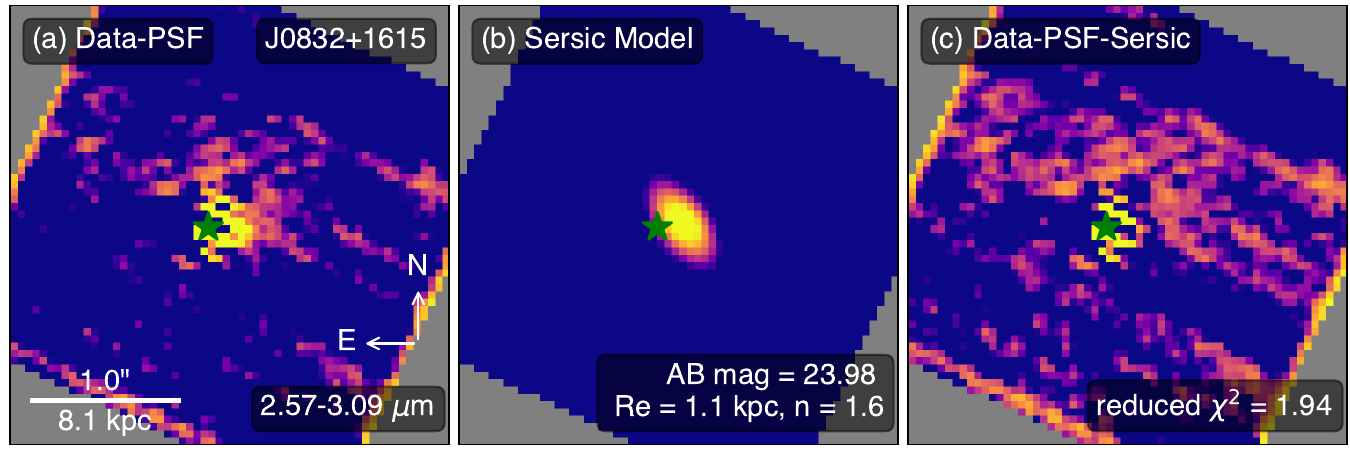}
\includegraphics[width=0.49\textwidth]{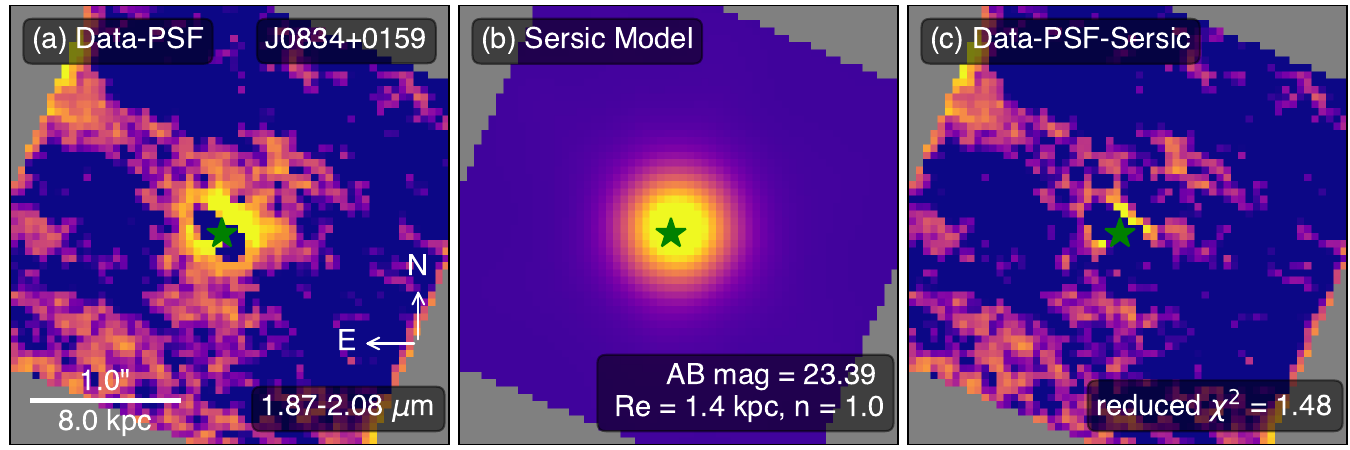}
\includegraphics[width=0.49\textwidth]{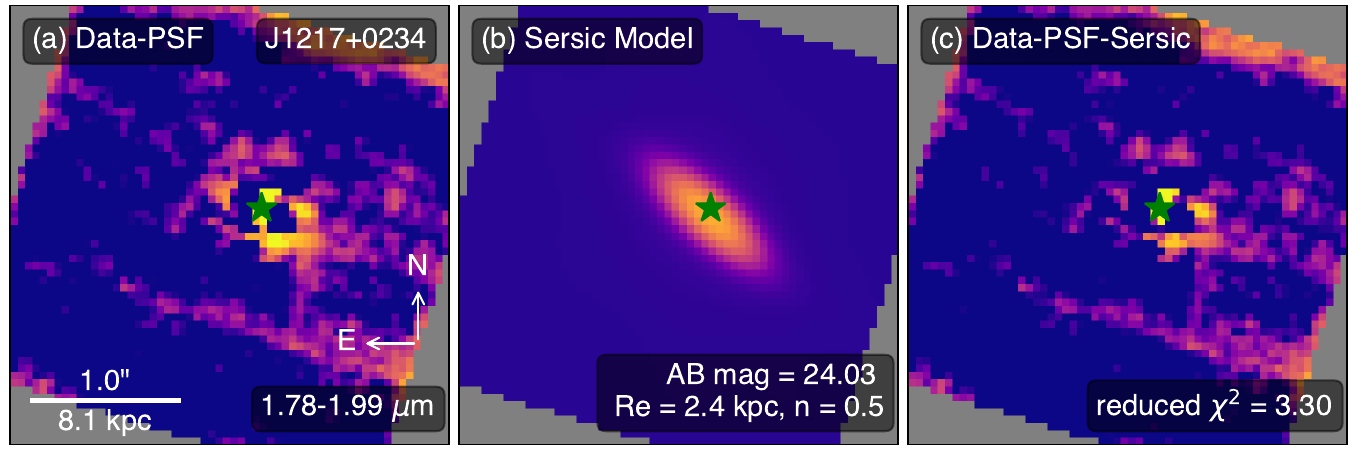}
\includegraphics[width=0.49\textwidth]{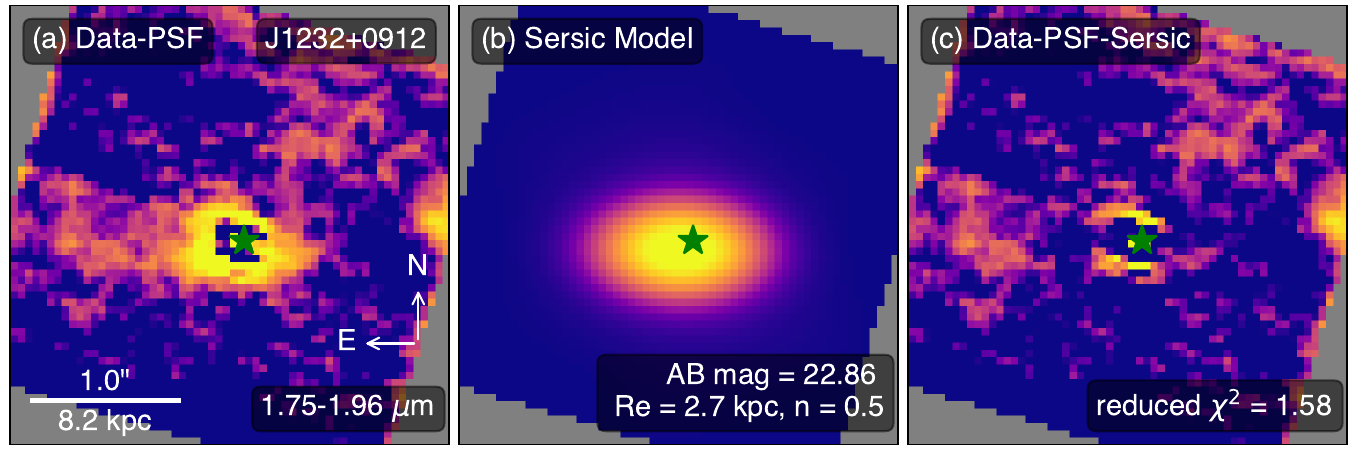}
\includegraphics[width=0.49\textwidth]{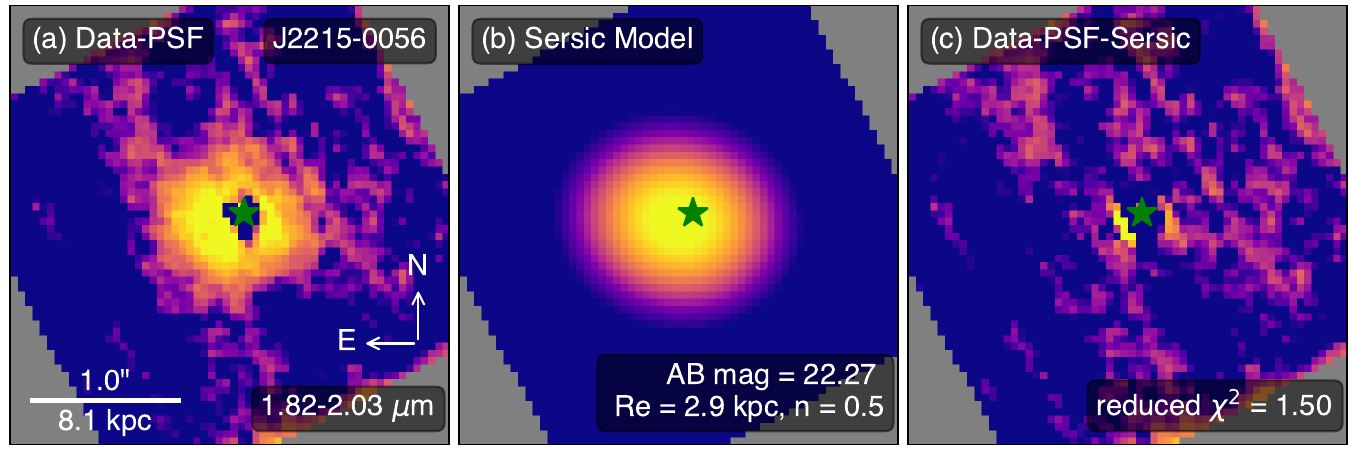}
\includegraphics[width=0.49\textwidth]{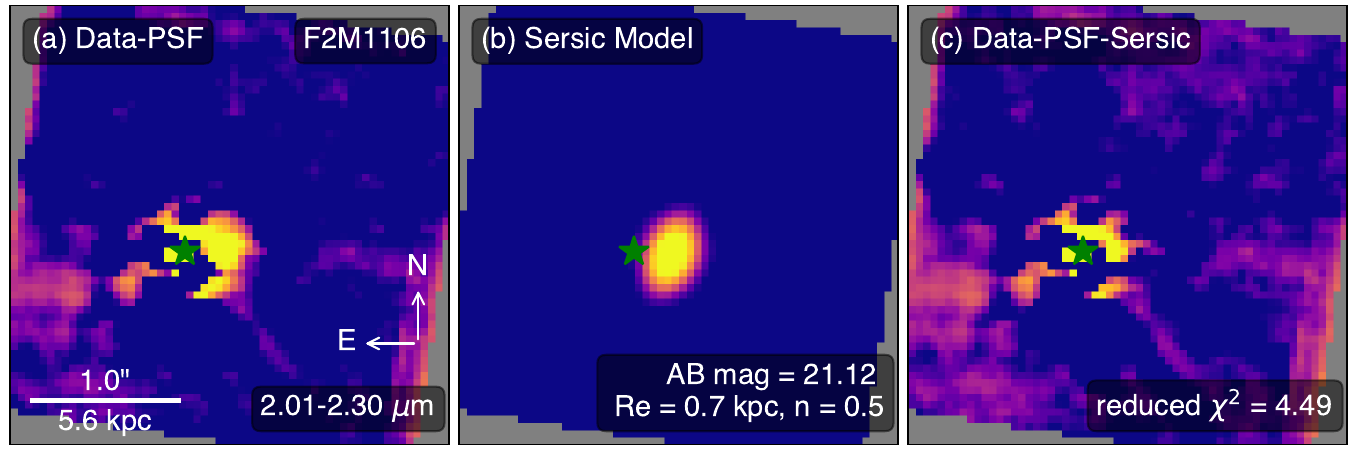}
\includegraphics[width=0.49\textwidth]{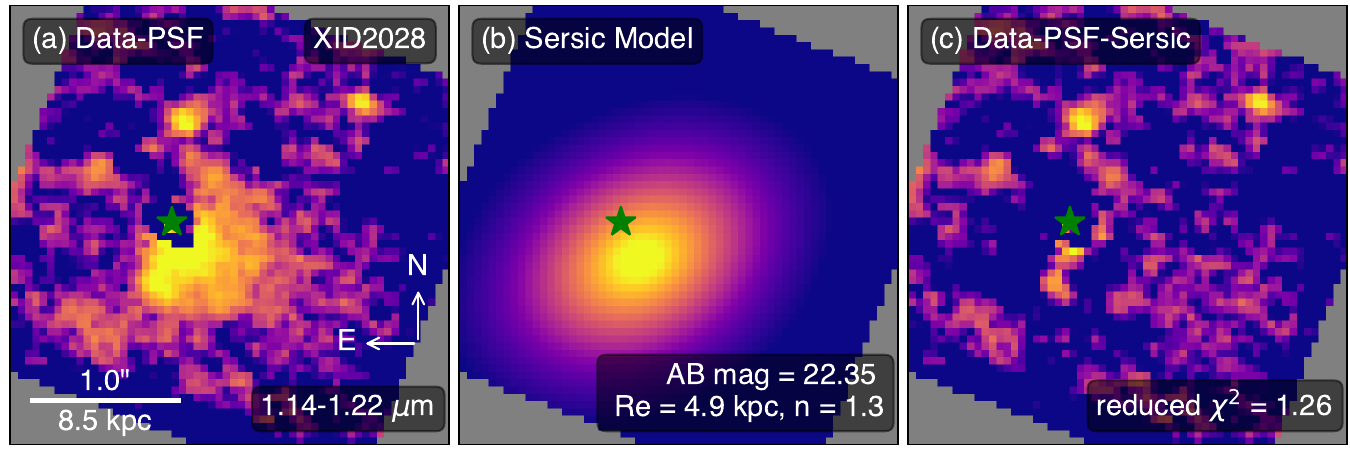}
\includegraphics[width=0.49\textwidth]{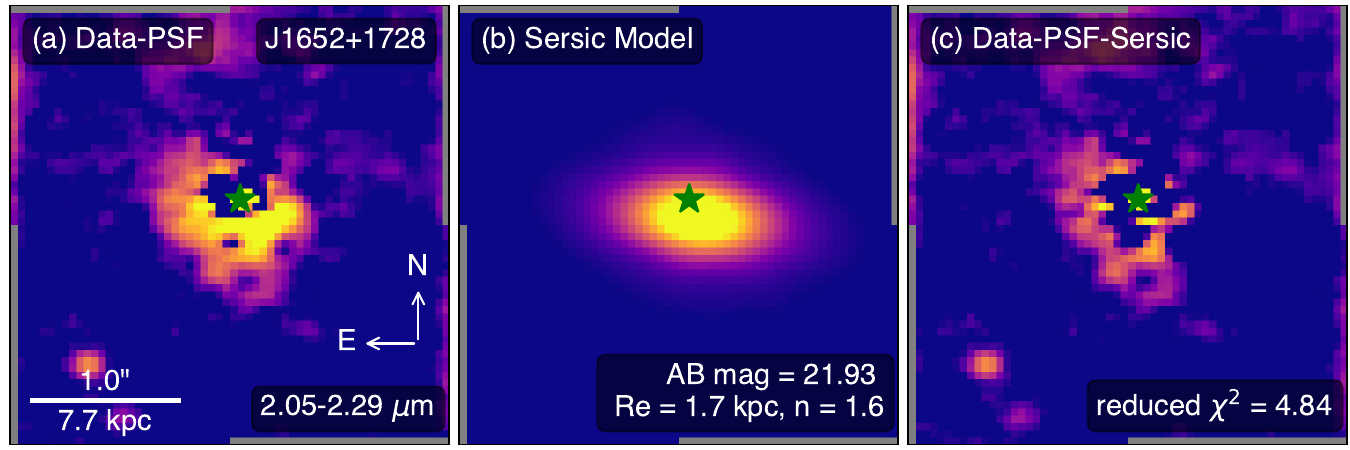}
    
    \caption{
    Two-dimensional host galaxy fitting results. We used \texttt{GALFIT} on wavelength-average PSF-subtracted images to model the host galaxy's profile. From left to right, the sub-panels display: (a) the wavelength-average PSF-subtracted image, (b) the best-fit Sersic profile, and (c) the corresponding residual image. The green stars mark the quasar positions. The north is up and the east to the left We found that J0832+1615, J1217+0234, and F2M1106 did not provide well-constrained fits.
    }
    \label{fig:galfit_output}
\end{figure*}

We derive the host galaxy morphologies for all targets as described in \autoref{sec:galfit_specfit}. The results of the \texttt{GALFIT} modeling are presented in \autoref{fig:galfit_output}, and the corresponding parameters, including magnitude, effective radius, and Sersic index, are summarized in \autoref{tab:galfit}. Due to residuals from imperfect PSF subtraction, the Sersic indices are often poorly constrained, with some values reaching the limits of the allowed fitting range. With the exception of J1217–0056, for which the centroid is fixed at the quasar position, all host galaxies show positional offsets from the quasars. We discuss possible explanations for these offsets in \autoref{dis:offcenter}. However, the Sersic fits for J0832+1615, J1217–0056, and F2M1106 are not well constrained, and their derived parameters should be interpreted with caution. Additionally, XID2028 and J1652+1728 exhibit irregular or clumpy structures that cannot be well modeled by a single Sérsic profile. Using the residual images, we estimated the light contributions from those clumps are less than 5 per cent of the host galaxy lights.

\begin{deluxetable*}{ccccccc}
 \tablecaption{Fitted host galaxy models using \texttt{GALFIT}.
 \label{tab:galfit}}
 \tablehead{ \colhead{Name} & \colhead{Observed $\lambda$} & \colhead{Restframe $\lambda$} & \colhead{Magnitude} & \colhead{R$_e$} & \colhead{n} & \colhead{PSF-to-Host}\\ 
 \colhead{} & \colhead{($\mu$m)} & \colhead{($\mu$m)} &\colhead{(AB)} & \colhead{(kpc)} &  } 
 \colnumbers
 \startdata
SDSSJ0832+1615 & 2.57-3.09 & 0.75-0.90 & 23.98$\pm$0.07$^?$ & 1.1$\pm$0.2$^?$ & 1.6$\pm$1.0$^?$ & 48$^?$\\
SDSSJ0834+0159 & 1.87-2.08 & 0.52-0.58 & 23.39$\pm$0.08 & 1.4$\pm$0.1 & 1.0$\pm$0.3 & 13\\
SDSSJ1217+0234 & 1.78-1.99 & 0.52-0.58 & 24.03$\pm$0.12$^?$ & 2.4$\pm$0.4$^?$ & 0.5$\pm$0.4$^?$ & 105$^?$\\
SDSSJ1232+0912 & 1.75-1.96 & 0.52-0.58 & 22.86$\pm$0.02 & 2.7$\pm$0.1 & 0.5$\pm$0.1 & 8.6\\
SDSSJ2215$-$0056 & 1.82-2.03 & 0.52-0.58 & 22.27$\pm$0.02 & 2.9$\pm$0.1 & 0.5$\pm$0.03 & 3.1\\
F2M1106 & 2.01-2.30 & 1.25-1.50 & 21.11$\pm$0.18$^?$ & 0.71$\pm$0.03$^?$ & 0.5$\pm$0.2$^?$ & 170$^?$\\
XID2028 & 1.14-1.22 & 0.44-0.47 & 22.34$\pm$0.07 & 4.9$\pm$0.3 & 1.3$\pm$0.1 & 3.2\\
SDSSJ1652+1728 & 2.05-2.29 & 0.52-0.58 & 21.93$\pm$0.01 & 1.70$\pm$0.04 & 1.6$\pm$0.1 & 26\\
\enddata
\tablecomments{Column 1: Target name. Column 2: Observed wavelength range ($\mu$m) used for the image. Column 3: Restframe wavelength range ($\mu$m). Column 4: Fitted AB magnitude of the host galaxy. Column 5: Effective radius in kpc. Column 6: Sérsic index. Column 7: Quasar to host galaxy flux ratio. The errors are 1$\sigma$ statistical uncertainties, which might be slightly underestimated. The measurements in the sources without a decent fit are marked with ``?".}
\end{deluxetable*}

\subsection{1D host spectra}

\begin{figure*}
  \centering
\includegraphics[width=0.49\textwidth]{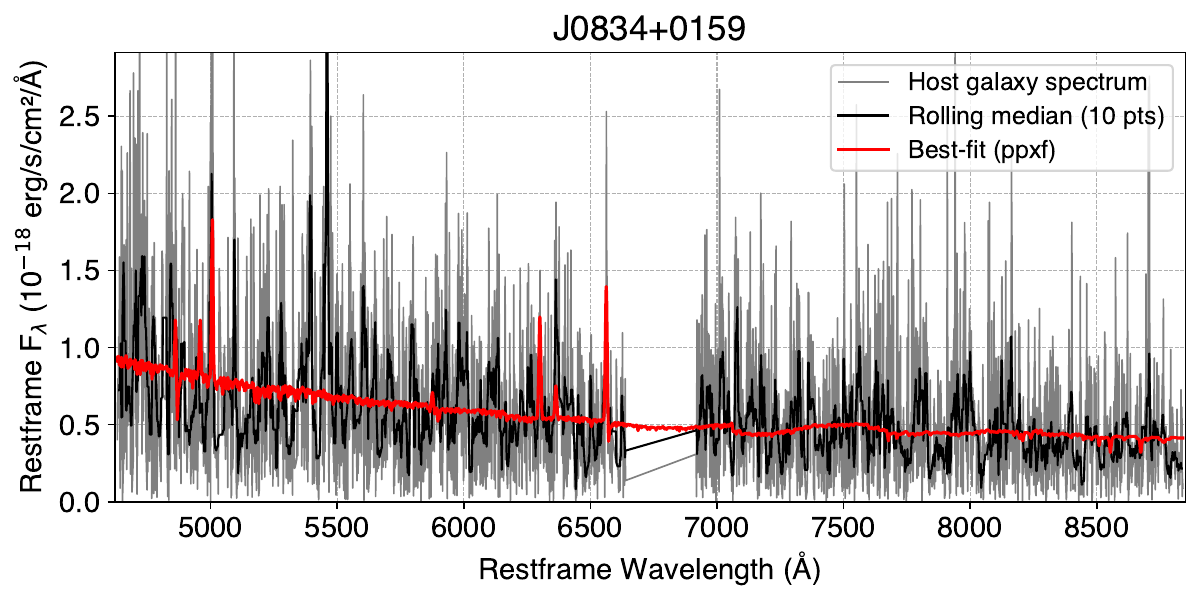}
\includegraphics[width=0.49\textwidth]{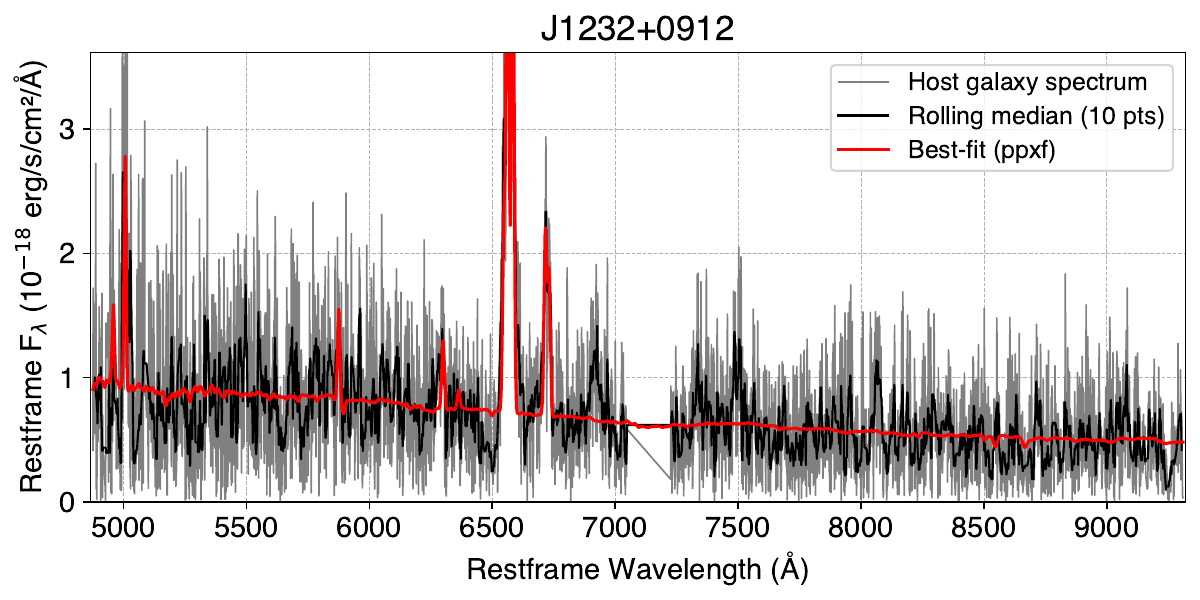}
\includegraphics[width=0.49\textwidth]{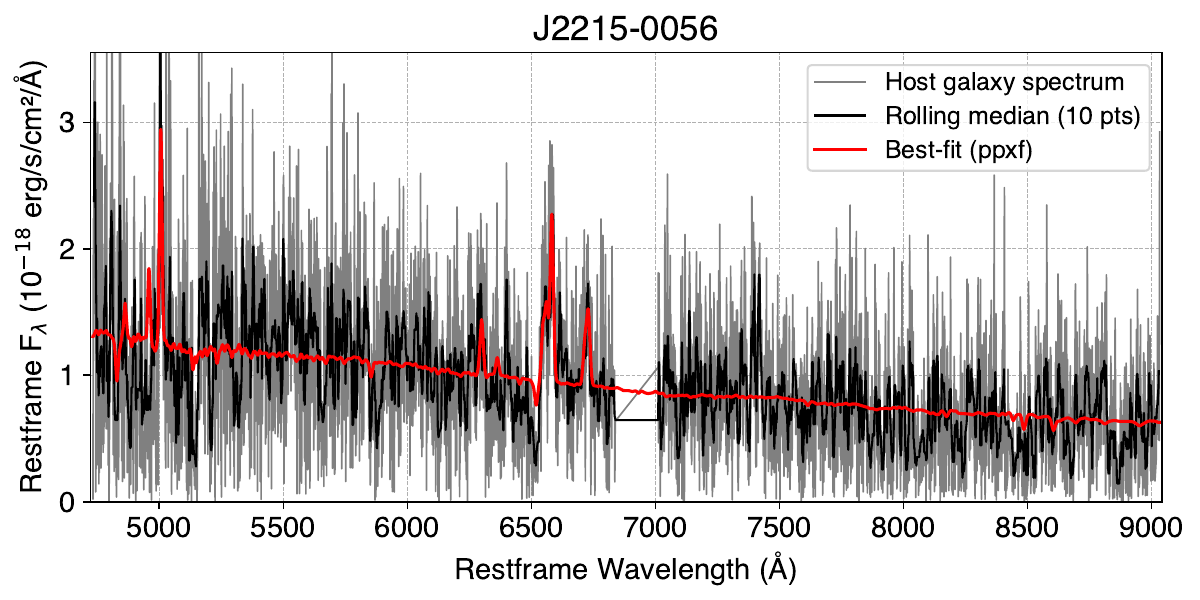}
\includegraphics[width=0.49\textwidth]{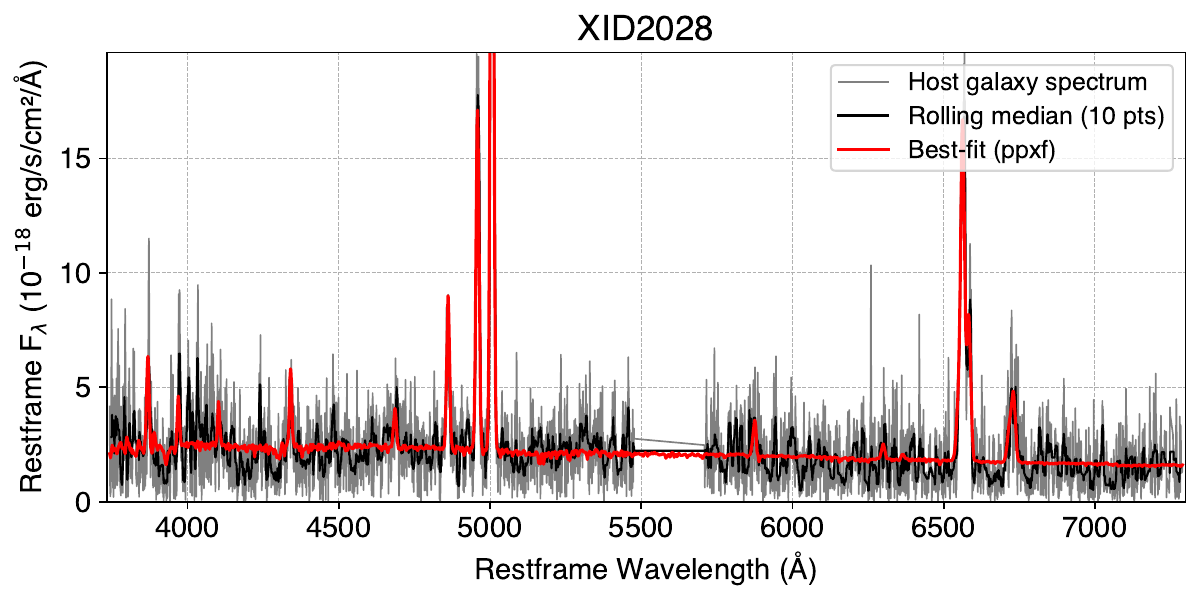}
\includegraphics[width=0.49\textwidth]{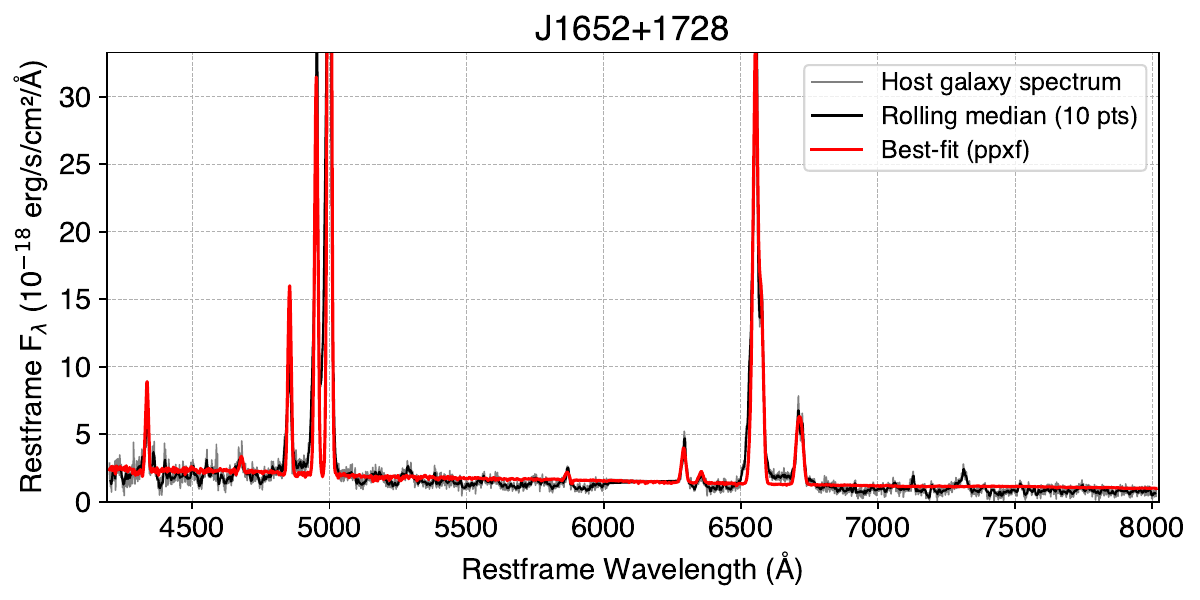}

    \caption{The 1-dimensional host galaxy spectra using the forced photometry with fitted host galaxy model. The black lines are data and the red lines are best-fit lines from \texttt{pPXF}
    }
    \label{fig:host_spectra}
\end{figure*}

We present the host galaxy spectra for five targets -- J0834+0159, J1232+0912, J2215$-$0056, XID2028, and J1652+1728 -- in \autoref{fig:host_spectra}. These are the five (out of eight) targets for which we can reliably isolate the host galaxy emission and use forced photometry to extract the one-dimensional host spectra. All spectra are free of significant spectral ``wiggles", with the exception of J1652+1728. Although our technique is less susceptible to wiggles than purely spectral decomposition methods, the host spectrum of J1652+1728 still shows weak wiggle patterns, likely due to its bright PSF (AB magnitude of 18.4 at 2$\mu$m) and a high PSF-to-host ratio of $\sim$26.
Using the extracted 1D spectra, we fit stellar population templates of varying ages and metallicities \citep{Bruzual2003} with the \texttt{pPXF} code \citep{Cappellari2017,Cappellari2023}, as detailed in \autoref{sec:ppxf_fit}. While some targets, such as J2215$-$0056, show hints of weak absorption lines, the low signal-to-noise ratios limit our ability to measure stellar velocity dispersions. Nonetheless, we successfully derive the stellar luminosities $\nu L_{\nu }$, mass-to-light ratios, and stellar masses $M_{\star}$ of the host galaxies. The best-fit values are summarized in \autoref{tab:ppxf}. 
The mass-to-light ratios are derived in the rest-frame Johnson V-band ($\sim$0.55$\mu m$). Stellar luminosities $\nu L_{\nu }$ are calculated by convolving the best-fit model spectrum with the Johnson V-band filter. The 1$\sigma$ uncertainties in luminosity and stellar mass incorporate systematic errors from various Sersic models (with indices of 0.5, 1, 2, and 4) and a fiducial 0.3 dex uncertainty in the mass-to-light ratios \citep{Conroy2009}.

\begin{deluxetable}{cccc}
 \tablecaption{ The derived host galaxy properties from \texttt{pPXF}.
 \label{tab:ppxf}}
 \tablehead{ \colhead{Name} & \colhead{log($\frac{\nu L_{\nu}}{L_\odot}$)} & \colhead{M-to-$L_V$ Ratio} & \colhead{log($\frac{M_\star}{M_\odot}$)} \\ 
 \colhead{} & \colhead{} & \colhead{(M$_\odot$/$L_\odot$)} & \colhead{(M$_\odot$)} } 
 \colnumbers
 \startdata
SDSSJ0834+0159 & 10.72$\pm$0.11 & 0.77  & 10.61$\pm$0.33 \\
SDSSJ1232+0912 & 10.75$\pm$0.12 & 0.77  & 10.64$\pm$0.31\\
SDSSJ2215$-$0056 & 10.94$\pm$0.11 & 0.84  & 10.87$\pm$0.32 \\
XID2028 & 10.70$\pm$0.13 & 1.68 & 10.92$\pm$0.32 \\
SDSSJ1652+1728 & 11.19$\pm$0.08 & 0.51 & 10.90$\pm$0.32 \\
\enddata
\tablecomments{Column 1: Target name. Column 2: Stellar luminosity of the host galaxy (L$_\odot$) in the restframe Johnson V-band ($\sim0.55$\um) filter (corresponding to observed wavelength of $\sim$1.4-2.1\um\ at $z$=1.6-2.9). Column 3: Stellar mass-to-light ratio in the restframe Johnson V-band filter. Column 4: Stellar mass. The 1$\sigma$ uncertainties account for systematic errors arising from different Sersic models and a fiducial 0.3 dex uncertainty in the mass-to-light ratios.}
\end{deluxetable}

\section{Discussion} \label{sec:discussion}

\subsection{Stellar mass and size-luminosity relation}

The stellar masses of our red and obscured quasars ($\sim$10$^{10.6-10.9}$ M$_{\odot}$) are consistent with the average stellar mass of quasar hosts as expected from abundance matching and clustering analysis. This consistency aligns with findings from the clustering and abundance matching of quasars at $z=1\sim3$, which suggest that they reside in haloes with masses $\sim$10$^{12-13}$ M$_{\odot}$ \citep{Eftekharzadeh2015,DiPompeo2015,Rodriguez-Torres2017}. This halo mass range is then translated into a corresponding stellar mass of $\sim$10$^{10-11}$ M$_{\odot}$ using the stellar mass–halo mass relation \citep{Behroozi2013}. These results confirm the robustness of our measurements and the link to these theoretical models in characterizing quasar host galaxies.

\begin{figure}
  \centering
  \includegraphics[width=0.9\columnwidth]{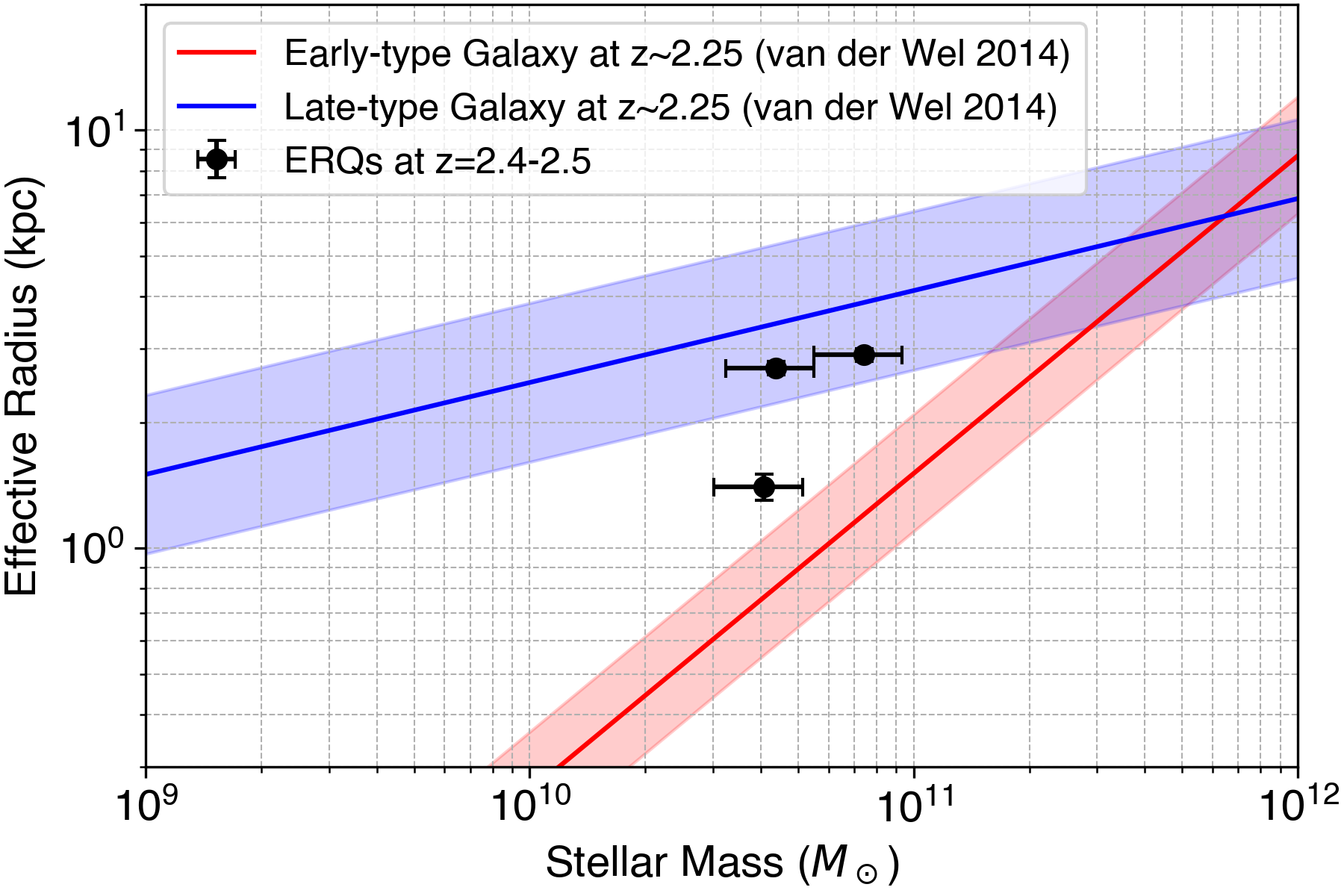}
    \caption{The effective radius as a function of stellar luminosity for the ERQs in comparison with inactive galaxy samples \citep{vanderWel2014}. The blue (red) line represents the relation for early-type (late-type) galaxies at $z\sim2.25$. The 1$\sigma$ errors of the relations are marked with color-shaded regions. 
    }
    \label{fig:re_stellarmass}
\end{figure}

Moreover, we compare the host galaxy morphologies of our obscured quasars to those of inactive galaxies. We focus on three ERQ targets (J0834+0159, J1232+0912, and J2215–0056), for which we obtained reliable single-component host galaxy fits using \texttt{GALFIT}, and which do not exhibit prominent tidal features (unlike, e.g., XID2028 and J1652+1728). In Figure \ref{fig:re_stellarmass}, we plot the half-light radii of the host galaxies as a function of stellar mass, in comparison with early-type and late-type galaxies at similar redshifts from \citet{vanderWel2014}. The three ERQ host galaxies lie between the size–mass relations for early-type and late-type galaxies, with two of them falling closer to the late-type population. This is consistent with their relatively blue colors and disk-like shapes (Sérsic indices $n \sim 0.5$–1). Their intermediate positions may suggest that these ERQ hosts are in a transitional phase from star-forming to quiescent galaxies, with quenching potentially driven by the extreme quasar outflows \citep{Zakamska2016_erq, Hamann2017}. However,the Sersic indices may be biased toward lower values due to reduced central flux from slight PSF over-subtraction. It is important to note that the size–mass relations in the comparison sample are based on inactive galaxies, and further study of host galaxy continua in unobscured quasars will allow for a more direct comparison.

\subsection{Why are red quasars off-centered?}\label{dis:offcenter}

Many of the obscured quasars and ERQs in our sample show significant (0.4-1.3 kpc) offsets between the centroids of the host galaxy continua and the quasars. Similar off-center morphology has also been seen in the ionized gas component such as the narrow \oiii\ line in XID2028 \citep{Veilleux2023} and J1652+1728 \citep{Vayner2024}, strengthening this off-center finding. We explore several possible explanations for these offsets. One possibility is that dust-reddened quasars, particularly ERQs, represent a transitional phase occurring at the final stage of a galaxy merger. During this phase, inflowing gas fuels quasar activity, while powerful outflows begin to expel dust and gas \citep{DiMatteo2005,hopkins08}. Although the young quasar remains embedded and obscured by dust, compact and fast outflows may have already formed \citep{Zakamska2014,Lau2024}.
The observed offsets could be linked to the merger process itself. Following the merger of two galaxies, the resulting non-axisymmetric gravitational potential can drag SMBHs toward the center via dynamical friction, eventually leading to coalescence \citep{begelman80,Yu2002}. The merged SMBH may then receive a recoil kick due to anisotropic gravitational wave emission, displacing it from the galactic center \citep{Loeb2007,Blecha2008}. Several of our targets, such as XID2028 and J1652+1728, display tidal features and possible galaxy companions, further supporting the post-merger scenario. 
Additionally, distant luminous quasars hosted by massive galaxies such as sub-millimeter galaxies and radio galaxies exhibit strong clustering and are suggested to reside in dense environments \citep{Brodwin2008,Chapman2009,Hickox2012,Wylezalek2013,Wylezalek2014}. One of our ERQs, J1652+1728, has also been observed to reside in an overdense region \citep{Wylezalek2022}. While a more robust statistical argument is needed to determine whether ERQs generally reside in overdense regions, mergers are more likely to occur in such environments.

Another possible explanation is non-uniform dust obscuration across the host galaxy, which can shift the fitted centroid of the host galaxy toward less obscured regions. \citet{Vayner2023} reported that for one ERQ in our sample, J1652+1728, the V-band extinction varies between 0.5 and 3 magnitudes in star-forming clumps located in the northeastern part of the host galaxy, based on elevated \ha/\hb\ ratios ($>3$). We observe a lack of stellar continuum in the same direction, consistent with the emission line analysis. Although the signal-to-noise ratio of the stellar continuum is too low on a per-spaxel basis to derive an extinction map, detailed emission line ratio analysis across the full sample of ERQ may confirm the partial obscuration hypothesis.

\subsection{Black hole mass - stellar mass scaling relation}

\begin{figure}
  \centering
\includegraphics[width=0.95\columnwidth]{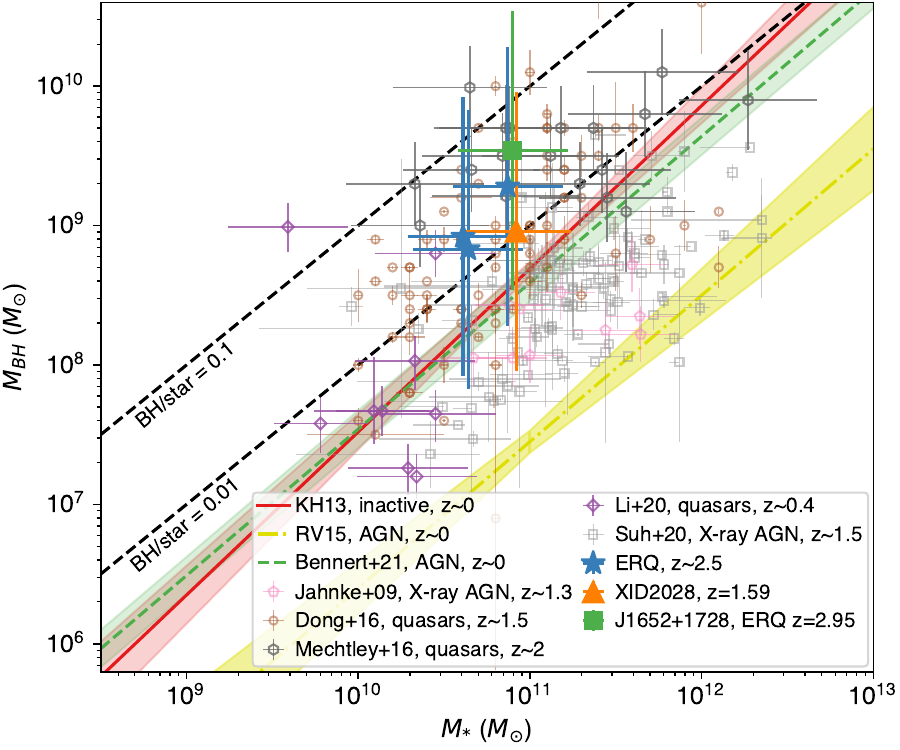}
    \caption{The black hole mass - stellar mass relation of the 5 ERQs and obscured quasars in comparison to various quasar samples and correlations \citep{KormendyHo2013,Reines2015,Bennert2021,Dong2016,Mechtley2016,Suh2020,Li2021}. The error bars and shaded regions represent 1$\sigma$ uncertainties.
    }
    \label{fig:Mbh-Mstellar}
\end{figure}

Studying the stellar mass–black hole mass (M\textsubscript{BH}–M\textsubscript{*}) relation provides critical insight into the co-evolution of galaxies and their central black holes \citep{KormendyHo2013}. Using stellar masses derived from host galaxy spectral fitting and black hole masses estimated from broad emission lines, we compare our obscured quasars to other quasar samples and established scaling relations. We extract nuclear spectra and measure the total luminosity and FWHM of the broad H$\alpha$ line. Details of the emission line measurements and analysis are presented in Neustadt et al. (in prep.). Black hole masses are estimated using the H$\alpha$ calibration from \citet{GreeneHo2005}: \begin{equation}
\frac{M_{\mathrm{BH}}}{M_{\odot}} = 2^{+0.4}_{-0.3} \times 10^6 
\left( \frac{L_{\mathrm{H}\alpha}}{10^{42}\ \mathrm{erg/s}} \right)^{0.55}
\left( \frac{\mathrm{FWHM}_{\mathrm{H}\alpha}}{10^3\ \mathrm{km/ s}} \right)^{2.06}.
\end{equation}
Single-epoch M\textsubscript{BH} estimates for ERQs can vary significantly, with a typical 1$\sigma$ uncertainty of $\sim$1 dex based on multi-line analyses \citep{Bertemes2025}. 
\autoref{fig:Mbh-Mstellar} shows the M\textsubscript{BH}–M\textsubscript{*} relation for the five targets with reliable stellar mass measurements, compared to various quasar samples and empirical scaling relations from both the local universe and at the redshift of our targets  \citep{KormendyHo2013,Reines2015,Bennert2021,Dong2016,Mechtley2016,Suh2020,Li2021}. Our targets lie in the upper-left region of the distribution, with M\textsubscript{BH}/M\textsubscript{*} ratios of 0.01–0.04. These values are elevated by $\sim$0.5–2 dex above the local relations \citep{KormendyHo2013, Reines2015, Bennert2021}.

HotDOGs, a population of heavily obscured quasars that are close relatives of ERQs, are also thought to be ``over-massive" for their hosts \citep{Eisenhardt2012, Wu2012}. Based on rest-frame optical to mid-infrared SED modeling of $\sim$100 HotDOGs at 
$z=1-4.6$, \citet{Assef2015} found that their black hole masses significantly exceed expectations from the local black hole–bulge mass relation \citep{Bennert2011}, even when only considering upper limits on stellar mass. The inferred black hole-to-spheroidal mass ratios of the HotDOGS are 
$\gtrsim0.01-0.2$, comparable in magnitude to those we find in our ERQ targets (0.01–0.04).

With the advent of JWST, studies in deep fields \citep{Eisenstein2023, Finkelstein2023} have revealed numerous high-redshift SMBH candidates that appear to be over-massive relative to their host galaxies \citep{Kocevski2023,Harikane2023,Greene2024,Maiolino2024}. While the non-AGN origin of these so-called ``little red dots" (LRDs) remains a possibility \citep{Baggen2024}, they exhibit several features—such as redder colors in rest-frame optical wavelengths, broad emission lines and over-massive features — that are similar to ERQs and HotDOGs. Though \citet{Stepney2024} found a heavily reddened quasar at cosmic noon shares remarkable similarities with LRD, whether obscured quasars are analogs of LRDs at cosmic noon cannot be concluded without a detailed comparison between the two populations, matched in properties such as luminosity and redshift. ERQs are primarily selected from SDSS, which constrains their redshift range to approximately 1 $\lesssim z \lesssim$ 3, whereas most LRDs are found at $z \gtrsim 4$. However, this redshift boundary may be somewhat arbitrary, driven largely by selection criteria and survey design (e.g., wavelength coverage). Large and deep sky surveys like Hyper Suprime-Cam and Euclid have recently identified LRD candidates at $z<4$ \citep{Ma2025,Bisigello2025}. These identifications will provide excellent comparison samples to investigate the connection between LRD and obscured quasars. It is also important to consider that the seemingly over-massive nature of LRDs or any luminosity-selected quasar samples may arise from selection biases inherent in flux-limited surveys \citep{Shen2015, Volonteri2023, Li2025}.

\subsection{Comparison with studies using HST and ground-based adaptive optics}

We compare our stellar continuum measurements with those from previous studies using HST and ground-based adaptive optics. \citet{Zakamska2019} analyzed HST 1.4–1.6\um\ images of 10 ERQs and successfully detected host galaxy continuum emission. Five of our ERQ targets are included in their sample, and for four of them (J0834+0159, J1232+0912, J2215–0056, J1652+1728), both half-light radius and optical luminosity measurements are available. The half-light radii reported by \citet{Zakamska2019} (3.0–6.5 kpc) are approximately twice as large as our measurements (1.4–2.9 kpc), likely due to the lower angular resolution of HST. The PSF FWHM of HST is about 0.15 arcseconds ($\sim$1.2 kpc at $z=2.4$) at 1.5 \um, which results in more blended quasar and host galaxy emission, compared with JWST. This blending can cause a loss of central flux and lead to biased estimates that favor more extended structures.

The optical luminosities we derive for the four shared targets ($10^{10.72-11.19} L_{\odot}$) are slightly lower (by $\sim$0–0.5 dex) than those reported in \citet{Zakamska2019} ($10^{10.66-11.76} L_{\odot}$), though consistent within the uncertainties. The lower stellar masses in our results may stem from more accurate morphological modeling enabled by JWST's higher resolution and reduced contamination from the quasar PSF. Additionally, despite careful filter selection, HST broadband imaging may still include light from extended emission lines (e.g., \oiii, \hb), leading to higher inferred luminosities. Lastly, the filters used differ slightly between the two studies: our measurements are based on V-band (0.54 \um) data, while \citet{Zakamska2019} used B-band (0.44\ um), which may also contribute to systematic differences.

\citet{Lau2024} conducted adaptive optics integral-field spectroscopic observations of 10 ERQs at $z\sim2.3-3.0$ using Keck/OSIRIS and Gemini/NIFS. Three of our targets, J1217+0234, J1232+0912 and J1652+1728, are included in their sample. While their study primarily focuses on extended emission lines, \citet{Lau2024} also investigate the presence of continuum emission. They apply two methods—the radial profile analysis and PSF subtraction—to assess whether the continuum is spatially extended. Among the three shared targets, they report that J1217+0234 and J1232+0912 are spatially resolved based on deviations from reference PSF profiles. However, due to limited wavelength coverage and data sensitivity, full PSF decomposition could not be performed, and thus no size or luminosity measurements are available for direct comparison with our results.



\section{Conclusions} \label{sec:conclusion}

In this paper, we present the host galaxy properties of six ERQs at $z=2.4-2.9$ and two dust-obscured quasars at $z=$ 0.4 and 1.6, based on JWST NIRSpec IFU data. We employ image decomposition techniques across the spectral range to successfully remove quasar emission. Using wavelength-averaged images, we model the host galaxy morphologies and derive key properties such as half-light radius and optical luminosity. For the five targets where the host galaxy emission can be reliably isolated, we extract the host spectra using forced photometry and perform full spectral fitting with stellar population templates.

We uncover several noteworthy findings listed below from our analysis of the targets, shedding light on their host galaxy properties, black hole growth, and potential evolutionary stages.

\begin{itemize}
    \item The stellar masses of five our red and obscured quasars ($\sim10^{10.6-10.9}$M$_{\odot}$) are consistent with theoretical predictions for quasar hosts derived from abundance matching and clustering analysis, validating our measurements and their link to models characterizing quasar host galaxies.
    \item The host galaxies of the three ERQs at $z \sim 2.4$ have half-light radii ranging from 1.4 to 2.9 kpc. Their sizes and stellar masses place them between the size–mass relations of early-type and late-type galaxies at similar redshifts, suggesting they may be in a transitional phase from star-forming to quiescent systems.
    \item 
    Most of our obscured quasars and ERQs exhibit significant spatial offsets between quasar and host galaxy centroids. Notably, XID2028 and J1652+1728 display prominent tidal features. These findings may result from post-merger dynamics, such as SMBH recoil and/or galaxy asymmetries in a transitional phase, and suggest that obscured quasars could be triggered by mergers. Alternatively, non-uniform dust obscuration may shift the apparent position of the host galaxy light.
    \item We find that ERQs lie 0.5–2 dex above the local M\textsubscript{BH}–M\textsubscript{*} relations, with black hole-to-stellar mass ratios of 0.01–0.04. Similar mass ratios have been observed in other heavily obscured quasars, such as the HotDOG population. This possible over-massive nature resemble those of JWST-identified SMBH candidates at higher redshifts. However, detailed comparisons between samples with matched properties is needed and selection biases should also be considered.
    \item Our measurements reveal more compact host galaxies and slightly lower optical luminosities compared to those reported using HST, likely due to JWST’s superior resolution and reduced contamination from quasar emission. We also find lower stellar masses, which may reflect more accurate morphological modeling and minimal line contamination in our IFU data.
\end{itemize}

In conclusion, the exceptional capabilities of JWST, particularly its high-resolution NIRSpec IFU observations, have proven invaluable in revealing the quasar host galaxies at cosmic noon. With the new techniques to separate quasar and host emissions, model host morphologies, and provide detailed spectral information, our results offer unparalleled insight into the evolutionary stages of quasars and their surrounding environments. As JWST observes more quasars, both obscured and unobscured, the increasing volume of IFU data on quasars at cosmic noon will yield a sizable sample for deriving a more statistically significant argument. Building on this, future 20-40 m ground-based telescopes equipped with adaptive optics -- such as the Extremely Large Telescope, Thirty Meter Telescope, and Giant Magellan Telescope --will offer angular resolution several times superior and sensitivity tens of times greater. This will enable the acquisition of high signal-to-noise host spectra and a detailed characterization of their host galaxy environments.  

\begin{acknowledgments}

This work is based on observations made with the NASA/ESA/CSA James Webb Space Telescope. The data were obtained from the Mikulski Archive for Space Telescopes at the Space Telescope Science Institute, which is operated by the Association of Universities for Research in Astronomy, Inc., under NASA contract NAS 5-03127 for \jwst. These observations are associated with program \#1335, \#2457, and \#3399.
Support for programs GO-02457, GO-03399 and ERS-01335 (YCC, NLZ, AV) was provided by NASA through grants from the Space Telescope Science Institute, which is operated by the Association of Universities for Research in Astronomy, Inc., under NASA contract NAS 5-03127.

\end{acknowledgments}

%

\vspace{5mm}
\facilities{JWST(NIRSpec)}


\software{numpy \citep{numpy}, astropy \citep{Astropy2013,Astropy2018,Astropy2022}, \texttt{pPXF} \citep{Cappellari2017,Cappellari2023}, \texttt{q3dfit} \citep{Rupke2014-ifsfit,Rupke2021-questfit}}





\bibliography{ref_new}{}
\bibliographystyle{aasjournal}



\end{document}